\documentclass[12pt]{article} 
\usepackage{psfig}
\usepackage{a4}

\newcommand{\bra}{\langle}
\newcommand{\ket}{\rangle}
\newcommand{\half}{\frac{1}{2}}

\newcommand{\gm}{\gamma}

\newcommand{\om}{\omega}  

\newcommand{\Gm}{\Gamma}

\newcommand{\be}{\begin{equation}}
\newcommand{\ee}{\end{equation}}
\newcommand{\bea}{\begin{eqnarray}}
\newcommand{\eea}{\end{eqnarray}}
\newcommand{\bean}{\begin{eqnarray*}}
\newcommand{\eean}{\end{eqnarray*}}
\newcommand{\nn}{\nonumber}
\newcommand{\veck}{{\mathbf k}}
\newcommand{\vecp}{{\mathbf p}}
\newcommand{\vecq}{{\mathbf q}}
\newcommand{\vecr}{{\mathbf r}}
\newcommand{\vecx}{{\mathbf x}}
\newcommand{\vecy}{{\mathbf y}}

\newcommand{\lesssim}{\raisebox{-.3ex}{$\enskip\stackrel{<}{\scriptstyle\sim
}\enskip$}}       
\newcommand{\ba}{\begin{eqnarray}}
\newcommand{\ea}{\end{eqnarray}}
\newcommand{\1}{q_1}
\newcommand{\2}{q_2}
\newcommand{\3}{q_3}
\newcommand{\4}{q_4}
\newcounter{saveeqn}

\begin{document}
\title{
\vskip -100pt
{  
\begin{normalsize}
\mbox{} \hfill \\
\mbox{} \hfill HD-THEP-00-32\\
\mbox{} \hfill hep-ph/0007357\\
\mbox{} \hfill July 2000\\
\vskip  70pt
\end{normalsize}
}
Exact and Truncated Dynamics in Nonequilibrium Field Theory
\author{
Gert Aarts,\thanks{email: aarts@thphys.uni-heidelberg.de}
\addtocounter{footnote}{1}
Gian Franco Bonini,\thanks{email: bonini@thphys.uni-heidelberg.de}
\addtocounter{footnote}{2} 
and 
Christof Wetterich\thanks{email: C.Wetterich@thphys.uni-heidelberg.de}
\addtocounter{footnote}{3}\\
\normalsize
{\em Institut f\"ur theoretische Physik, Universit\"at Heidelberg}\\
\normalsize
{\em Philosophenweg 16, 69120 Heidelberg, Germany}\\
\normalsize
}
}
\date{July 31, 2000}
\maketitle
 
\renewcommand{\abstractname}{\normalsize Abstract} 
\begin{abstract}
\normalsize 

Nonperturbative dynamics of quantum fields out of equilibrium is often
described by the time evolution of a hierarchy of correlation functions,
using approximation methods such as Hartree, large $N_f$, and
$n$PI-effective action techniques. These truncation schemes can be
implemented equally well in a classical statistical system, where results
can be tested by comparison with the complete nonlinear evolution obtained
by numerical methods. For a $1+1$ dimensional scalar field we find that
the early-time behaviour is reproduced qualitatively by the Hartree
dynamics. The inclusion of direct scattering improves this to the
quantitative level. We show that the emergence of nonthermal temperature
profiles at intermediate times can be understood in terms of
the fixed points of the evolution equations in the Hartree approximation.
The form of the profile depends explicitly on the initial ensemble. While
the truncated evolution equations do not seem to be able to get away from
the fixed point, the full nonlinear evolution shows thermalization with a
(surprisingly) slow relaxation.

\end{abstract}

\newpage

\section{Introduction}

An understanding of the dynamical evolution of nonequilibrium
quantum fields is needed in diverse physical situations as
nonrelativistic condensed matter, relativistic heavy-ion colliders, and
the early universe. 
For several reasons a theoretical description is expected to be difficult.
First, effective irreversibility has to arise from time-reversal invariant
equations. Second, for (asymptotically) late times one expects a
nonequilibrium system to thermalize, which implies an effective
independence of the initial state, and prescribes a definite value for all
correlation functions. Furthermore, during the nonequilibrium evolution,
the answer to the question: what dominates the full dynamics at a certain
stage, might itself be time-dependent. When approximation methods are
used, the chosen method will often need to be modified or replaced as time
goes on, to incorporate this shift in importance.

To illuminate this, consider as a prototype for a nonequilibrium system
the universe at the end of inflation (see e.g.\
\cite{Kofman:1997yn,Boyanovsky:1996sv}). In this case several distinct
regimes are easily identified. While the early stage is dominated by the
oscillating inflaton field, in the intermediate regime the created quanta
scatter both with the inflaton and with each other, leading to a partial
energy redistribution between the modes. Finally, in the last stage
subsequent interactions are expected to bring the universe to thermal
equilibrium. It is a great theoretical challenge to describe these various
stages in a unified way.

Although in principle the time evolution of expectation values is
determined completely, after the initial density matrix is given, by the
microscopic Heisenberg equations of motion, this is in practice only of
minor help, due to the absence of exact solutions or solution methods.
Consider for example a simple scalar quantum field theory with a quartic
interaction. The Heisenberg equation of motion determines the time
evolution of the mean field, 
\be
(\partial_t^2 - \nabla_x^2 +m^2)\bra\phi(x,t)\ket =
-\lambda\bra\phi^3(x,t)\ket/2,
\ee
where the brackets indicate the expectation value with respect to the
initial density matrix. 
This equation requires knowledge of the three-point
function $\bra\phi^3(x,t)\ket$. The equation for the three-point
function itself involves the five-point function (in general $n+2$-point
functions are needed to solve the exact evolution for $n$-point
functions), which leads to a full hierarchy of
coupled equations. 
In most cases a solution to this hierarchy is not available, and 
it becomes of interest to find approximation schemes that capture the
essential part of the full dynamics as correctly as possible. 

A widely-used approach is to truncate the infinite hierarchy of
correlation functions. One of the simplest truncations is the Hartree
approximation in which at most two-point functions appear, and the
three-point function is replaced by $\bra\phi^3(x,t)\ket =
3\bra\phi(x,t)\ket\bra\phi^2(x,t)\ket$. A systematic way to implement this
is by using a large $N_f$ expansion, where $N_f$ denotes the number of
e.g.\ scalar or fermion fields 
\cite{Cooper:1994hr} (for applications, see e.g.\
\cite{Cooper:1997ii,Boyanovsky:1997cr,Baacke:1997se,Destri:1999hd}). 
The main drawback of these Gaussian truncations is that in homogeneous or
translationally invariant ensembles scattering is absent, which limits the
range of
validity. This feature has been an important stimulation to improve upon
the homogeneous Hartree and large $N_f$ approximations. One possibility is
to allow for inhomogeneous mean fields. In that case scattering via the
space-dependent mean field $\bra \phi(x,t)\ket$ is present
in the effective equations (see e.g.\ 
\cite{Boyanovsky:1996zy} for analytical investigations and
\cite{Aarts:1998td} for a numerical study). 
Another natural extension is to go beyond the Gaussian approximation and
include higher nontrivial correlation functions 
\cite{Cheetham:1996nd,Mihaila:1997gb,Braghin:1998yw}, 
guided e.g.\ by the large $N_f$ expansion to next-to-leading order.
This typically results in effective equations that are nonlocal in time. 
{}From a practical (numerical) point of view, it is
desirable to use effective equations that are local in time, as in the
Hartree approximation. 
This has motivated the use of the 1PI-effective action for equal-time
correlation functions
\cite{Wetterich:1997ap,Bettencourt:1998nf,Bettencourt:1998xb}. 
Truncated at quartic order this includes scattering and all $1/N_f$
corrections \cite{Bonini:1999dh}. 
A discussion of the structure of dynamical equations for equal-time
correlation functions in the large $N_f$ expansion up to and
including $1/N_f^2$ terms can be found in \cite{Ryzhov:2000fy}. 
Finally, very promising results have been obtained recently
\cite{Berges:2000ur}, using a truncation of the 2PI-effective action at
next-to-leading order.  The resulting equations are nonlocal in
time, i.e.\ they require the integration of memory kernels. For a $1+1$
dimensional field theory this appears, however, not to be a numerical
obstacle.

In all cases, it would be desirable to perform a direct test of these
methods and their range of validity. Unfortunately a comparison with the
full nonperturbative evolution in the quantum field theory cannot be made,
due to the absence of exact methods. In the case of quantum mechanics on
the other hand, such tests can be performed, and the evolution from a
Hartree factorization, a $1/N_f$ expansion at leading and next-to-leading
order, and other extensions beyond the Gaussian aproximation have been
compared with the exact evolution obtained by numerically solving the
Schr\"odinger equation
\cite{Cheetham:1996nd,Mihaila:1997gb}.  It
is, however, not obvious how the lessons learned from quantum mechanics
with one degree of freedom can be translated to field theory with (in
principle infinitely) many degrees of freedom. Both scattering and the
possibility of taking the thermodynamic limit are absent in the quantum
mechanical case.

The situation in classical field theory is completely different.
Here the full evolution can be simulated using Monte Carlo methods
and numerical integration: initial conditions are generated by sampling 
according to the initial probability distribution, and
the subsequent time evolution follows from solving the classical equations
of motion, which can be done numerically. Expectation values are then
constructed by summing over many independent realizations. When the number
of initial conditions is taken larger and larger, the initial probability
distribution is approximated better and better, and the resulting time
evolution will become exact, in principle. 
As we will see below, it is possible to implement many of the
approximation methods discussed above for the quantum field theoretical
case also in a classical field theory. The reason is that the methods
do not directly touch upon the quantum nature, but instead state how to
truncate a hierarchy of correlation functions, which is present in a
classical statistical system as well. 
Therefore, we focus in this paper on nonequilibrium evolution in classical
field theory, formulated on a spatial lattice to regularize the theory. 
Note that the role of the thermodynamic limit may be investigated
keeping the lattice spacing fixed.
The general strategy is to compare the truncated dynamics with the
numerical results obtained from sampling initial conditions from a given
initial probability distribution. Our hope is that the insights obtained
below will survive when going to the quantum field theory.

In the remainder of the Introduction we give the outline of the paper. As
a simple toy model we consider a classical scalar field theory in
$1+1$ dimensions, with the (continuum) action
\be
\label{eqaction}
S = \int dtdx\, \left[\half (\partial_t\phi)^2 - \half (\partial_x
\phi)^2 -\half m^2\phi^2 -\frac{\lambda}{8}\phi^4\right].
\ee
In the next section we show how the (homogeneous) Hartree approximation
can be implemented in the classical theory. In Sec.\ \ref{seceffective} we
discuss evolution equations obtained from a functional differential
equation for the equal-time 1PI effective action.  A truncation of the
time-dependent effective action at quadratic order gives evolution
equations that are equivalent to those obtained in the Hartree
approximation.  Scattering is incorporated by a
truncation of the effective action that includes momentum-dependent
four-point functions. In Sec.\ \ref{secensemble} the choice of initial
probability distribution and some numerical and lattice aspects are
discussed. We have divided the comparison between the truncated evolution
equations and the lattice results in two parts. In Sec.\ \ref{secearly} we
discuss  the early and intermediate times, where the system is still
relatively far from thermal equilibrium. 
The late-time regime, where thermal equilibrium is approached, is
described in Sec.\ \ref{seclate}.
Our findings are summarized in Sec.\ \ref{secoutlook}.

\section{Classical Hartree approximation}

The classical problem is fully specified by the equation of motion
\be
\label{eqeq}
\partial_t^2\phi(x,t) = \left[\partial^2_x-m^2\right]\phi(x,t)-\lambda
\phi^3(x,t)/2,
\ee
supplemented with initial conditions for $\phi(x,t)$ and 
$\pi(x,t)=\partial_t\phi(x,t)$. These initial conditions are determined by
the initial probability distribution $\rho[\pi(x),\phi(x)]$, where
$\phi(x)=\phi(x,0), \pi(x)=\pi(x,0)$. 
The choice of initial distribution is not needed at this stage, we only 
assume that the ensemble is translationally  invariant in space and
respects the discrete symmetry $\phi\to -\phi$, $\pi\to -\pi$.
The average with respect to the initial distribution will be denoted with
brackets
$\bra\cdot\ket$.
Our aim is to find evolution equations for (equal-time) correlation
functions of the field $\phi(x,t)$ and the canonical momentum $\pi(x,t)$.

As discussed in the Introduction, a widely-used approach in quantum field
theory is the Hartree approximation. This can be implemented in the
classical theory as well, and a brute but simple way to do this
is as follows. If we assume a Hartree-type factorization for the
interaction term in the classical equation, i.e., $\lambda \phi^3 \to
3\lambda \phi\bra\phi^2\ket$, we find
\be
\label{eqh}
\partial_t^2\phi(x,t) = \left[\partial^2_x-m^2-\frac{3}{2}\lambda\bra
\phi^2(x,t)\ket\right]\phi(x,t).
\ee
In the case of translationally invariant ensembles, $\bra\phi^2(x,t)\ket$
is independent of $x$ and the equation can be written in momentum space
as 
\be
\label{eqhartree}
\partial_t^2\phi(q,t) = -\bar\om_q^2\phi(q,t),
\ee
with the effective frequency squared
\be
\label{eqfreq}
\bar\om_q^2=\om_q^2+\frac{3}{2}\lambda \bra\phi^2\ket,
\ee 
where $\om_q^2=q^2+m^2$, and $\bra\phi^2\ket = 
\bra\phi^2(x,t)\ket$. In the case of $N_f$ scalar fields with a
complete $O(N_f)$ symmetry, the second term on the right-hand-side of
Eq.\ (\ref{eqfreq}) is multiplied by $(N_f+2)/(3N_f)$. In this paper we
restrict ourselves to the case $N_f=1$. Note, however, that there is no
problem in principle to extend the analysis below to finite $N_f>1$.
The unequal-time two-point function
\be
S(x-y;t,t') = \bra \phi(x,t)\phi(y,t')\ket = 
\int \frac{dq}{2\pi}\, e^{iq(x-y)}S(q;t,t'),
\ee  
obeys in this approximation the usual mean-field equation of motion
\be
\left[\partial_t^2+\bar\om_q^2\right]S(q;t,t')=0,
\ee
where the effective frequency is determined from the two-point function
at equal time:
\be
\bar\om_q^2=\om_q^2+\frac{3}{2}\lambda \int
\frac{dp}{2\pi}\,S(p;t,t).
\ee
Note that modes with a given momentum $q$ only interact with the
homogeneous mean-field background, so that direct scattering between
different momentum modes is absent.

The dynamics can be written equivalently\footnote{In the sense that the
evolution equations presented below are obtained from the same starting
point (\ref{eqhartree}).} in terms of equal-time
expectation values, at the expense of introducing more than one 
two-point function. The following four combinations are a priori
independent:
\bea
\nn
&& G_{\phi\phi}(x-y,t) = \bra\phi(x,t)\phi(y,t)\ket,\\
\label{eqtwo}
&&G_{\pi\pi}(x-y,t) = \bra\pi(x,t)\pi(y,t)\ket,\\ 
\nn
&&G_{\pi\phi}(x-y,t) = \half\bra\pi(x,t)\phi(y,t)+\phi(x,t)\pi(y,t)\ket,
\eea 
and the parity-odd combination 
\be
G^{\rm odd}_{\pi\phi}(x-y,t) =
\bra\pi(x,t)\phi(y,t)-\phi(x,t)\pi(y,t)\ket.
\ee 
The dynamical equations are conveniently written in momentum space,
according to
\be
G_{\phi\phi}(q,t) = \int dx\, e^{-iqx}G_{\phi\phi}(x,t),
\ee
etc., and they couple only the combinations (\ref{eqtwo}):
\bea 
\nn &&\partial_t G_{\phi\phi}(q,t) = 2G_{\pi\phi}(q,t),\\
\label{eqLO}
&&\partial_t G_{\pi\phi}(q,t) = -\bar\om_q^2 G_{\phi\phi}(q,t) +
G_{\pi\pi}(q,t),\\ 
\nn 
&&\partial_t G_{\pi\pi}(q,t) = -2\bar\om_q^2
G_{\pi\phi}(q,t). 
\eea 
The fourth combination $G^{\rm odd}_{\pi\phi}(q)$ is exactly conserved
under the Hartree equations and does not enter in the dynamics. It is zero
in the case that the ensemble is invariant under space
reflection.

In the Hartree approximation a nontrivial combination of
two-point functions, termed $\alpha^2$ in \cite{Bettencourt:1998nf},
is conserved, 
\be 
\label{eqalpha}
\alpha^{-2}(q) = G_{\phi\phi}(q,t)G_{\pi\pi}(q,t) - G^2_{\pi\phi}(q,t),
\ee
for each $q$.
This can be understood from the absence of scattering or mode mixing 
at this order and the resulting symmetry (see Appendix \ref{app}).
Finally, the Hartree equations conserve the expectation value of the 
energy, 
\be
E_{\rm Hartree} = L\int \frac{dq}{2\pi}\,\left[ 
\half G_{\pi\pi}(q,t) + 
\half \left(
\om_q^2 + \frac{3\lambda}{4} \int \frac{dp}{2\pi}\,G_{\phi\phi}(p,t)
\right)
G_{\phi\phi}(q,t)\right],
\ee
which can be obtained using a Gaussian factorization of the $\phi^4$-term
in the microscopic expression for the energy, or from the effective
Lagrangian given in Appendix \ref{app}.

\section{Equal-time effective action}
\label{seceffective} 

In order to improve upon the Hartree approximation, it is necessary to
include direct scattering contributions in the evolution equations. We aim
here at a formulation that is local in time. This is also desirable from
a numerical point of view.\footnote{See however \cite{Berges:2000ur}
for a successful implementation of time-nonlocal evolution equations.}
One way to achieve this is by employing a formalism based on the
equal-time effective action $\Gamma [\phi,\pi;t]$, the generating
functional of 1PI equal-time correlation functions
\cite{Wetterich:1997ap}. 
The effective action obeys the following (functional) evolution equation
\cite{Bettencourt:1998nf,Bonini:1999dh}
\be
\partial_t \Gamma[\phi, \pi;t] = - {\cal L}_{\rm cl} 
\Gamma[\phi, \pi;t],
\label{e1}
\ee
with
\bea
\nn
 {\cal L}_{\rm cl}  &=& \int dx\, \left[ \pi(x) 
{\delta \over \delta \phi(x)} \right. 
+ \phi(x)\left(\partial_x^2 - m^2
- \half\lambda\left[\phi^2(x) + 3 \bar G_{\phi\phi}(x,x)\right] 
\right)\frac{\delta}{\delta \pi(x)} \nonumber
\\ 
&&  -  
\int dx_1  dx_2  dx_3\, \bar G_{\phi \psi_1}(x,x_1)
\bar G_{\phi \psi_2}(x,x_2)\bar G_{\phi \psi_3}(x,x_3) \nonumber\\ 
&&\left.\times {\delta^3 \Gamma \over 
\delta \psi_1(x_1) \delta \psi_2(x_2) \delta \psi_3(x_3)}  
{\delta \over \delta \pi(x) } \right].
\eea
Here $\psi \equiv (\phi,\pi)$ and $\bar G_{\psi \psi'}(x,x')$ denotes
the full (matrix) propagator in arbitrary field background, obtained from
$\Gm$
as
\be
\label{eqGGm}
\bar G^{-1}_{\psi\psi'}(x,y)=
\frac{\delta^2\Gamma}{\delta\psi(x)\psi'(y)}.
\ee
The full propagator evaluated at zero background is written without the
bar.
The effective action depends on `effective' fields $\phi$ and $\pi$, which 
are defined via the Legendre transformation, relating $\Gm$ and
$\ln Z$ in the usual way \cite{Wetterich:1997ap}. Though we use the same
notation, these fields should not be confused with the microscopic fields
that appear in the original action (\ref{eqaction}).

In order to solve the exact evolution equation (\ref{e1}), some
approximation has to be made. This brings us back to the issue of
truncations, as discussed in the previous sections. 
We use a truncation or ansatz that includes all 1PI $n$-point functions,
with $n\leq 4$, and respects the symmetry $\psi\to-\psi$ as well
as spatial translation and reflection. The ansatz reads
\bea
&&\Gamma [\phi,\pi;t] = \int_q 
\left[  \half A(q)\phi^*(q)\phi(q)  + \half B(q) \pi^*(q) \pi(q) 
 +  C(q) \pi^*(q) \phi (q) \right] 
\nonumber \\ 
&&\;\;\;\; 
 + \frac{1}{8} \int_{q_1,q_2,q_3,q_4} 
2\pi\delta(q_1+q_2+q_3+q_4) \Big[
u(q_1,q_2,q_3)  \phi(q_1) \phi(q_2) \phi(q_3) \phi(q_4)   \nonumber \\ 
&& \;\;\;\;
+ v(q_1,q_2,q_3) \pi(q_1) \phi(q_2) \phi(q_3) \phi(q_4) 
+ w(q_1,q_2,q_3)\pi(q_1) \pi(q_2) \phi(q_3) \phi(q_4)
\nonumber \\
&& \;\;\;\;
+ y(q_1,q_2,q_3) \pi(q_1) \pi(q_2) \pi(q_3) \phi(q_4) 
+  z(q_1,q_2,q_3) \pi(q_1) \pi(q_2) \pi(q_3)  \pi(q_4) \Big]. 
\nonumber 
\eea
We use the shorthand 
\[
\int_q = \int \frac{dq}{2\pi},
\] 
and suppress the time dependence of the two-point couplings $A, B,
C$, the four-point couplings $u$, $v$, $w$, $y$, $z$, and the correlation
functions  $G_{\psi\psi'}$ in this section.

The matrix relation (\ref{eqGGm}) at vanishing background can now be given
explicitly, and
\bea
\nn
&&G_{\phi\phi}(q) = B(q)/\alpha^2,\\
&&G_{\pi\phi}(q) =  -C(q)/\alpha^2,\\ 
\nn
&&G_{\pi\pi}(q) = A(q)/\alpha^2,
\eea 
with the determinant 
\be 
\alpha^2(q) = A(q)B(q)-C^2(q).
\ee
This definition of $\alpha(q)$ is equivalent to that in Eq.\
(\ref{eqalpha}).
For future convenience we introduce 
\be
c(q) \equiv \frac{C(q)}{B(q)},
\ee
which will be used to convert two-point functions:
\be
G_{\pi\phi}(q) = -c(q)G_{\phi\phi}(q).
\ee

\begin{figure} \centerline{\psfig{figure=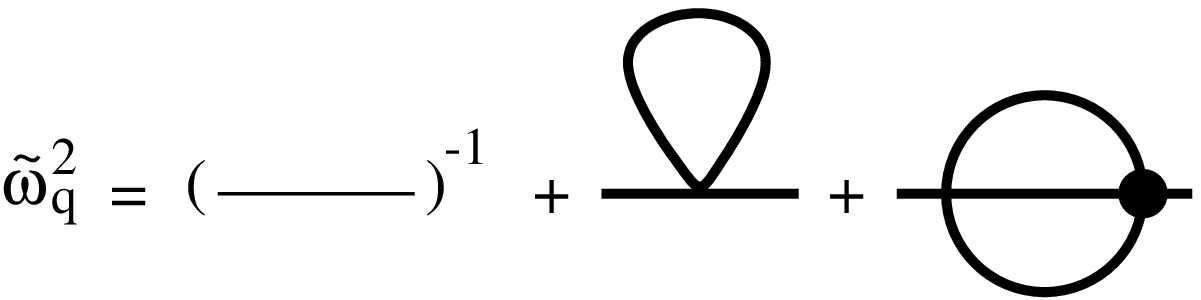,height=2.2cm}}
\caption{Graphical representation of the effective frequency squared, Eq.\
(\ref{eqfreq2}). The thick (thin) lines denote the full (free) equal-time
two-point function $G_{\phi\phi}$. The blob in the setting-sun diagram is
a full equal-time vertex function. 
} 
\label{figdiagram} 
\end{figure}

The time dependence of the effective action determined by (\ref{e1})
translates into evolution equations for the couplings. Exact flow
equations for the two-point functions follow from taking the second
derivatives of Eq.\ (\ref{e1}) with respect to $\phi$ and $\pi$ at
$\phi=\pi=0$:\footnote{The equations presented below are slightly simpler
than the ones that can be obtained from setting $N_f=1$ in the equations
for the $O(N_f)$ model \cite{Bonini:1999dh}.}
\ba
&&\partial_t A(q) = 2\tilde\omega^2_qC(q) \nonumber\\
&&\partial_t B(q) = -2C(q)-2\gamma(q)B(q) \label{fleqfirst}\\
&&\partial_t C(q) = -A(q)+\tilde\omega^2_qB(q) -\gamma(q) C(q),\nonumber
\ea
with the frequency squared
\bea
\label{eqfreq2}
\tilde\omega^2_q &=& \omega^2_q +
\frac{3\lambda}{2}\int_p\,G_{\phi\phi}(p) \\
\nonumber
&& -\frac{3\lambda}{8}\int_{q_1,q_2,q_3}2\pi\delta(q-q_1-q_2-q_3)\, 
G_{\phi\phi}(q_1)G_{\phi\phi}(q_2)G_{\phi\phi}(q_3) \Big[ 
4u(q_1,q_2,q_3) \\
\nonumber
&&  -3c(q_1)v(q_1,q_2,q_3)
 + 2c(q_1)c(q_2)w(q_1,q_2,q_3) -c(q_1)c(q_2)c(q_3)y(q_1,q_2,q_3)
\Big],
\eea
where we recognize the free part, the Hartree term, and a contribution
with the topology of the setting-sun diagram, containing three full
propagators and a (complicated) dynamical vertex function (see Fig.\
\ref{figdiagram}).
The other factor appearing in (\ref{fleqfirst}) has a setting-sun
structure as well:
\bea
\nonumber
\gamma(q) &=& \frac{3\lambda}{8}
\int_{q_1,q_2,q_3}2\pi\delta(q-q_1-q_2-q_3)\, 
G_{\phi\phi}(q_1)G_{\phi\phi}(q_2)G_{\phi\phi}(q_3)
\Big[  
v(q,-q_1,-q_2) \\
\nonumber
&& -2c(q_1)w(-q_1,q,-q_2) +c(q_2)c(q_3)\{y(q_3,q_2,-q)+2y(-q,q_1,q_3)\} \\
&& - 4c(q_1)c(q_2)c(q_3)z(-q_1,q,-q_2)
\Big].
\eea
These equations have to be completed with the evolution equations for the
four-point couplings, and those are listed in Appendix \ref{appfourpoint}.
Note that $\alpha(q)$ is no longer a conserved quantity as it is in the
Hartree approximation, but obeys $\partial_t\alpha(q) = -\gm(q)\alpha(q)$. 

Finally, the evolution equations conserve exactly the energy
\bea
E_\Gm &=& E_{\rm Hartree} -\frac{3}{8}\lambda 
\int_{q_1,q_2,q_3,q_4}2\pi\delta(q_1+q_2+q_3+q_4)\, G_{\phi\phi}(q_1) 
G_{\phi\phi}(q_2) \nonumber \\&&
\times G_{\phi\phi}(q_3) G_{\phi\phi}(q_4) \Big[ u(\1,\2,\3)-
c(\1)v(\1,\2,\3) 
+c(\1)c(\2)w(\1,\2,\3) 
\nonumber\\ &&
- c(\1)c(\2)c(\3)y(\1,\2,\3) + c(\1)c(\2)c(\3)c(\4)z(\1,\2,\3) 
\Big].
\label{eq:energy}
\eea

It is illuminating to make a connection with the Hartree equations derived
in the preceding section.  A truncation of the effective action at
quadratic order gives dynamical equations involving only $A,
B$, and $C$. It is straightforward to check that these give precisely the
Hartree equations (\ref{eqLO}) for the equal-time two-point functions.

The equal-time effective action permits an easy inclusion of quantum
effects \cite{Wetterich:1997rp}. The Hartree approximation does not
distinguish between classical and quantum field theories. In the quartic
truncation the quantum effects add simple terms to the evolution equations
of the quartic couplings \cite{Bettencourt:1998nf}. A direct verification
of the truncated evolution for quantum fields is obviously much harder.

\section{Initial ensemble and lattice discretization}
\label{secensemble} 

For an investigation of the time evolution of correlation functions we
need to specify the initial ensemble or probability distribution. In this
paper we choose to start from a Gaussian, translationally invariant
ensemble. The reasons to take an initially Gaussian ensemble are the
following: first of all, the Hartree equations truncate the dynamics to
Gaussian dynamics for all times. Therefore, possible truncation effects
will show up during the time evolution only, but not already initially,
since the initial ensemble is treated correctly in the Hartree
approximation. Furthermore, Gaussian ensembles are the ones that are often
considered in {\em quantum} field theory away from equilibrium. It is
straightforward to construct an initial density matrix that leads to
Gaussian correlation functions, and vice versa. Finally,
from a technical point of view, Gaussian ensembles can easily be
implemented in both the truncated evolution equations and the numerical
evolution obtained by sampling initial conditions.

In the set of Gaussian ensembles, we choose to take the equilibrium
distribution function of the unperturbed Hamiltonian $H_0$, where 
\bea 
\nn 
H &=& \int dx\,\left[
\half\pi^2 + \half (\partial_x \phi)^2 +\half m^2\phi^2
+\frac{\lambda}{8}\phi^4
\right]\\ 
&=& H_0+V, \;\;\;\;\;\;\;\;\ V=\int dx\,\frac{\lambda}{8}\phi^4. 
\eea 
Thus, the initial probability distribution is given by 
\be 
\rho[\pi(x),\phi(x)] = Z_0^{-1} \exp [-H_0/T_0], \;\;\;\;  
Z_0 = \int {\cal D}\pi{\cal D}\phi \, \exp [-H_0/T_0], 
\ee 
where $\phi(x)=\phi(x,0), \pi(x)=\pi(x,0)$, and 
\be 
\int{\cal D}\pi{\cal D}\phi = \int \prod_x d\pi(x) d\phi(x) 
\ee 
denotes the integral over the initial phase-space. The temperature of the
initial ensemble is denoted with $T_0$.  Note that this distribution
function is of course not the equilibrium distribution for nonzero
$\lambda$.

Since the ensemble is Gaussian, the only nontrivial correlation functions
at $t=0$ are the two-point functions, and they read
\bea
\nn
\bra \phi(q)\phi(q')\ket = G_{\phi\phi}(q,0)2\pi\delta(q+q'), 
&&\;\;\;\;\; G_{\phi\phi}(q,0) = T_0/(q^2+m^2),\\
\bra \pi(q)\pi(q')\ket = G_{\pi\pi}(q,0)2\pi\delta(q+q'),
&&\;\;\;\;\; G_{\pi\pi}(q,0)= T_0.
\label{eqgaussian}
\eea
Possible variations of this initial ensemble would be to choose
different `initial temperatures' $T_0(q)$ for each momentum mode and the
$\phi$ and $\pi$ fields.

To properly define the model, we formulate it on a lattice in space with
spacing $a$.  The number of spatial sites is $N$, such that the volume
is $L=Na$, and we use periodic boundary conditions. 
Due to the finite volume and lattice cutoff, the momentum $q$ takes a
finite number of discrete values: 
\be 
q=\frac{2\pi k}{L},
\;\;\;\;k=\left\{-\frac{N}{2}+1, \ldots,
\frac{N}{2}\right\},
\ee
and momentum integrals are replaced by sums:
$\int dq/(2\pi) \to L^{-1}\sum_q$.
The dispersion relation is modified due to the Laplacian on the lattice
and reads
\be  
\om^2_q= \hat q^2+m^2, \;\;\;\;\hat q^2 = \frac{2}{a^2}(1-\cos aq).
\ee
Classical field theory suffers from the Rayleigh-Jeans divergence, which
implies that the lattice cutoff cannot be taken to zero in a
straightforward manner. 
Indeed, the expectation value of the full energy in this ensemble is
\be
\label{eqfullenergy}
\bra H\ket = \bra H_0\ket +\bra V\ket = 
L\left[\frac{T_0}{a} + \frac{3\lambda}{8}\bra\phi^2\ket^2\right]\\
= N\left[T_0 + a\frac{3\lambda}{8}\bra\phi^2\ket^2\right],
\ee
where the first explicit expression is written such that
the extensivity of the energy and the linear divergence as $a\to 0$ are
manifest, and the second one on the other hand emphasizes `classical
equipartition' in a system with $N$ degrees of freedom.
In this paper we work at fixed lattice cutoff $ma=0.25$ (corresponding to  
a fixed momentum cutoff $\Lambda=\pi/a=4\pi m$). 
The thermodynamic limit can be taken by increasing the number of lattice
sites $N$, keeping the initial temperature $T_0$ fixed.

The expectation value $\bra\phi^2\ket$ can be calculated at $t=0$:
\be
\bra\phi^2(x)\ket = \frac{1}{L}\sum_q G_{\phi\phi}(q,0) = 
\frac{1}{L}\sum_q \frac{T_0}{\hat q^2+m^2}
\equiv \frac{T_0}{m}I.
\ee
The sum is ultraviolet finite. For the lattice sizes we use its value is
close to the infinite volume value: $I = \frac{1}{2}[1+a^2m^2/4]^{-1/2}
\simeq 0.5$.

The initial conditions for the evolution equations can now be given
explicitly. They read
\bea
\nn
&&A(q,0) = G^{-1}_{\phi\phi}(q,0) = (\hat q^2+m^2)/T_0,\\
&&B(q,0) = G^{-1}_{\pi\pi}(q,0) = 1/T_0,
\eea
and 
\bea
\nn
C(q,0) &=& u(q_1,q_2,q_3;0) = v(q_1,q_2,q_3;0) \\
&=& w(q_1,q_2,q_3;0) = y(q_1,q_2,q_3;0) = z(q_1,q_2,q_3;0) = 0.
\eea
The evolution equations are solved using a standard fourth-order
Runge-Kutta algorithm that is exactly time reversible.

The full nonlinear evolution is constructed by sampling initial conditions
from the Gaussian ensemble (\ref{eqgaussian}) and solving the equation of
motion numerically for each initial condition in real space. In order to
do this, the action is discretized on a lattice in time as well, with
step size $a_0<a$. The resulting discretized equations of motion are of
the
leap-frog type. We have used $a_0/a=0.05$ and $0.01$, and checked 
independence of the step size.
The ensemble is approximated by using many independent initial conditions.
The number of ensemble members is denoted with $N_m$. For the results
presented below we have used $N_m \sim 2,\!500-16,\!000$. 

\begin{figure}[t]
\centerline{\psfig{figure=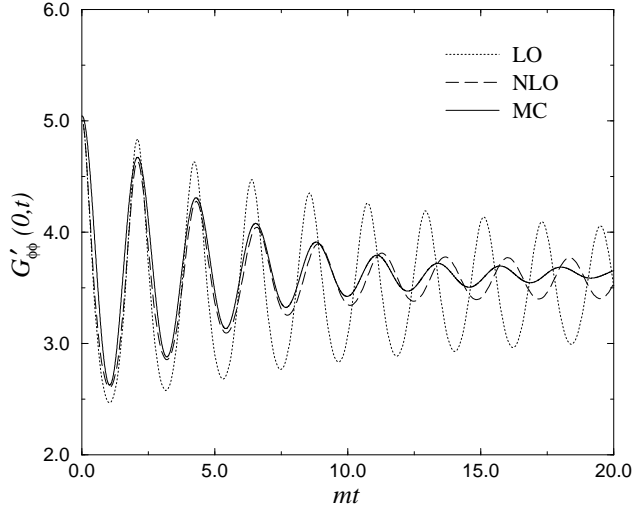,height=7cm}}
\caption{Time evolution of the two-point function $G'_{\phi\phi}(0,t)$. We
compare the Hartree equations (LO, $N=512$), the next-to-leading order
equations (NLO, $N=160$), and the results from a Monte Carlo sampling
($N=512, N_m=16,\!000$). In Figs.\ \ref{figGphi0}-\ref{figTprofilezoom} 
the initial temperature of the Gaussian ensemble is $T'_0=5$. In all
figures, the lattice spacing is $ma=0.25$. 
}
\label{figGphi0}
\end{figure}  

Finally, for the numerical analysis it is convenient to use the mass
parameter $m$ as the dimensionful scale. Therefore we rescale all
dimensionful parameters with the appropriate power of $m$. Rescaled
variables will be denoted with a prime. Furthermore, the classical
equation of motion (\ref{eqeq}) can be made independent of the coupling
$\lambda$ by introducing a field $\phi'$ as 
\be
\phi'=\sqrt{3\lambda/m^2}\,\phi.
\ee
The rescaled (dimensionless) canonical momentum is 
$\pi'=(3\lambda/m^4)^{1/2}\pi$ and the dimensionless energy reads
$E' = 3\lambda E/m^3$. We define a dimensionless temperature $T' =
3\lambda T/m^3$ such that $E/T= E'/T'$. As a result, besides the lattice
parameters $a'=ma$ and $N$ ($L'=a'N$), only one parameter remains to be
specified: 
\be
T'_0 \equiv \frac{3\lambda}{m^3}T_0.
\ee
A larger value of $T_0'$ corresponds to a larger effective interaction
strength.\footnote{A quick way to change to primed variables is to put
$m=1, \lambda=1/3$.}

\section{Early and intermediate times}
\label{secearly}

We are now fully equipped to compare the time evolution using the
different
methods. For shortness, we refer below to the evolution obtained in the
Hartree approximation as LO (leading order) and the evolution from the
equal-time effective action truncated up to four-point couplings as NLO
(next-to-leading order).  Although in principle there is no small
parameter governing the truncation, we use the labeling common in the
large $N_f$ expansion, since the Hartree approximation is closely related
to the leading-order contribution in $1/N_f$. The full nonlinear evolution
of a sample of initial conditions is denoted with MC (Monte Carlo).

The equal-time two-point function at zero momentum $G_{\phi\phi}(q=0,t)$
is shown in Fig.\ \ref{figGphi0}, for a typical choice of parameters. The
numerical integration of the NLO
evolution is rather time consuming, due to the presence of the four-point
couplings $u$, $v$, $w$, $y$, and $z$ that depend on three independent
momentum variables. Therefore it is not possible to take as large volumes
as in LO and MC. However, we have checked that at this stage this does not
affect the comparison. It is clear that the first few
oscillations are well approximated by both LO and NLO dynamics. We
will refer to this period as the early-time regime. It is visible that the
Hartree evolution underestimates damping whereas the size of the
oscillations in NLO remains comparable with the MC result much
longer. Around $mt=10$ the periods of oscillation in NLO and MC evolution
start to disagree.

\begin{figure}
\centerline{\psfig{figure=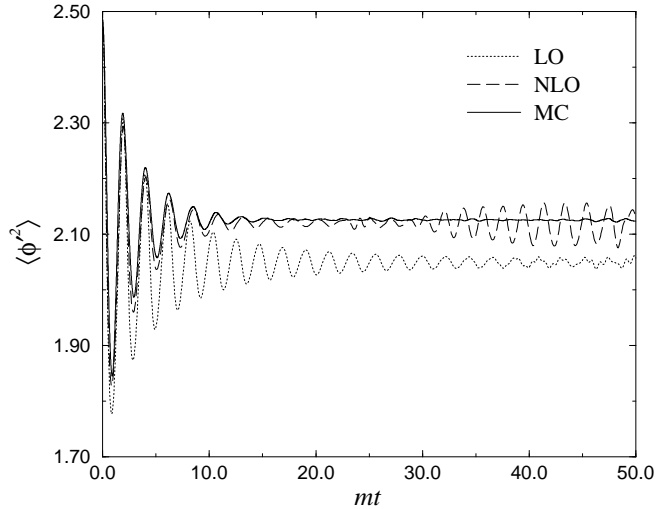,height=7cm}}
\caption{Time evolution of $\bra \phi'^2(x,t)\ket = L'^{-1}\sum_q 
G'_{\phi\phi}(q,t)$. The initial value is $\bra \phi'^2\ket = T_0'I =2.5$.
The time-averaged values in the interval $20<mt<50$ are 2.049 (LO), 2.114
(NLO), and 2.126(5) (MC). The Hartree fixed-point value (see below) is
2.056.
}
\label{figGav_50}
\end{figure}  

The crucial quantity in the mean-field approximation (\ref{eqh}) is the
field squared $\bra\phi^2(x,t)\ket = L^{-1}\sum_q
G_{\phi\phi}(q,t)$ which is presented in Fig.\ \ref{figGav_50}. Again we see
that the first few oscillations are in good agreement. Then a difference
becomes visible between LO on the one hand, and NLO and MC on the other hand:
at LO the amplitude of oscillations reduces slower (less damping) and the
averaged value lies below the other two. A comparison of the
time-averaged values between $20<mt<50$ shows that the Hartree result
differs by a few percent from NLO and MC. The time-averaged value in NLO is 
surprisingly close to the MC result.

\begin{figure} 
\centerline{\psfig{figure=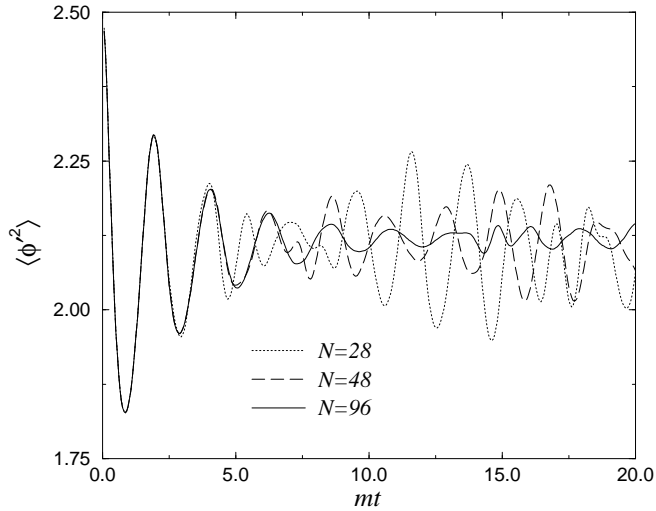,height=7cm}}
\caption{Volume dependence: NLO evolution of $\bra\phi'^2\ket$. In a 
smaller volume the evolution deviates earlier from the large-volume limit.
Damping can be seen only in the period before the evolution starts 
to deviate.
}
\label{figconvergence} 
\end{figure}

\begin{figure} 
\centerline{\psfig{figure=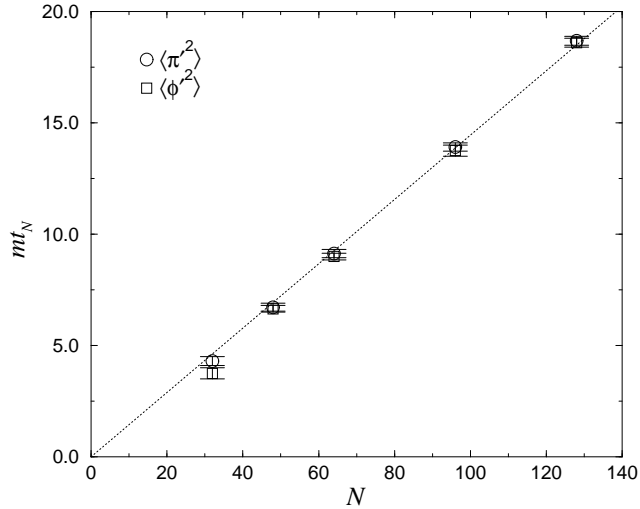,height=7cm}}
\caption{Thermodynamic limit in the NLO evolution: volume dependence of
the time $mt_N$ where $\Delta O_N(t)$ exceeds 0.005 (see text), for
$O=\pi'^2(x,t)$ and $\phi'^2(x,t)$. The dashed line is a straight-line fit
through the origin.
}
\label{figbreaktime} 
\end{figure}

After the initial reduction the size of fluctuations increases again in
NLO around $mt=30$. To investigate this issue, we study the volume dependence 
of the evolution in NLO. The phenomenon of `resurgence of fluctuations'
 is presented in Fig.\ \ref{figconvergence} for
relatively small volumes. We see that fluctuations become large again
after the initial reduction and that the evolution becomes undamped.
The time when this occurs increases with the volume. We stress
that this effect is not physical, as it is absent in the MC result, nor is
it generated by errors in the numerical integration. Rather, it is a
property of the NLO approximation. Since this seems to be an important 
limitation for the validity of NLO, we analyse the
thermodynamic limit and the consistency of the NLO evolution on a
quantitative level. We compare the dynamics for different volume sizes
with the largest one that is available ($N=160$). For an equal-time
observable $O(t)$, we define the difference
\be 
\Delta O_N(t) =
\frac{\left|\bra O_N(t)\ket -\bra O_{160}(t)\ket\right|}
{\bra O_{160}\ket_{\rm av}}. 
\ee 
The normalization $\bra O_{160}(t)\ket_{\rm av}$ is
the time-averaged value of $\bra O_{160}(t)\ket$ between $0<mt<20$, and is
used to set the scale. We denote the time where $\Delta O_N(t)$ exceeds
the conservative bound of $0.005$ by $t_N$. The dependence of $t_N$
on the volume is shown in Fig.\ \ref{figbreaktime}. It turns out that
$t_N$ is sensitive to the details of the evolution, due to the oscillating
character of $\Delta O_N(t)$. This sensitivity is indicated with error
bars. We see that in a larger volume the evolution behaves better for a
longer time, and that this time increases roughly linearly with the system
size.  We emphasize that the possibility of taking the thermodynamic limit
distinguishes the comparison performed in this paper with those where
quantum mechanics was used to test the truncated evolution. For this
aspect classical fields are closer to quantum fields than quantum
mechanics is.

At a later moment, to which we will refer as $t_{\rm u}$, fluctuations
grow rapidly and the numerical evolution becomes uncontrolled. Typically
$t_{\rm u}$ is much larger than $t_N$. We have checked that the time where
the evolution starts to behave badly is not an artefact of the numerical
integration. For instance, reducing the step size by a factor of 5 does
not affect the results. A similar behaviour has been noted before in a
system of anharmonic oscillators in $0+1$ dimensions
\cite{Bettencourt:1998xb}. The breakdown of the evolution equations beyond
leading order at large times has also been observed for quantum mechanics
\cite{Mihaila:1997gb}.

\subsection*{Fixed points in the Hartree approximation}

{}From the viewpoint of thermalization, the most interesting observable
is $G_{\pi\pi}(q,t)$. In an interacting theory the equilibrium value is
$G^{\rm eq}_{\pi\pi}(q) = T$ for all $q$. Away from equilibrium we define
therefore an `effective temperature' for a momentum mode $q$:
\be
T(q,t)\equiv G_{\pi\pi}(q,t),
\ee 
and the average temperature over all modes
\be
T(t) = N^{-1}\sum_q  G_{\pi\pi}(q,t) = a\bra\pi^2\ket.
\ee
Because of our choice of initial ensemble all momentum modes have
initially the same (noninteracting) temperature $T_0$, so that
$G_{\pi\pi}(q,0)$ is flat in momentum space. Due to the nonzero coupling
it deviates from being flat immediately after $t=0$. For later times
deviation from `flatness' of $G_{\pi\pi}(q,t)$ is a good measure for
deviation from thermal equilibrium.

\begin{figure} 
\centerline{\psfig{figure=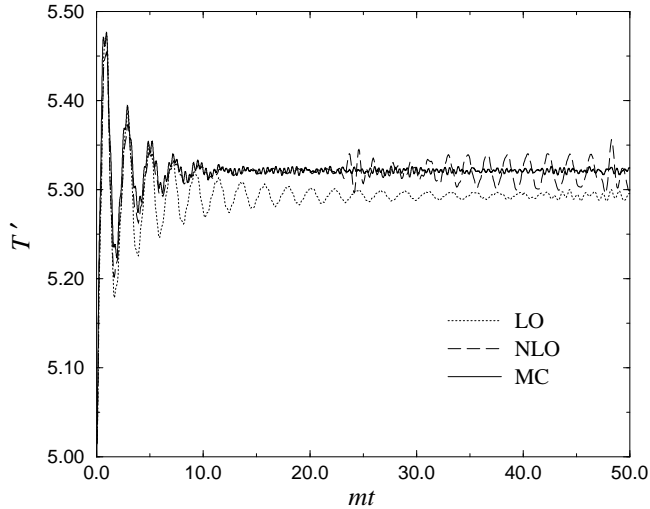,height=7cm}}
\caption{Time evolution of the effective temperature $T'(t)= N^{-1}\sum_q
G'_{\pi\pi}(q,t)$. The time-averaged values in the interval $20<mt<50$ are
5.293 (LO), 5.318 (NLO), and 5.321(4) (MC). The value at the Hartree
fixed point is 5.291.
}
\label{figTav_50} 
\end{figure}

In Fig.\ \ref{figTav_50} we show the average temperature $T'$ as a
function of time. The qualitative aspects comparing LO, NLO, and MC are
the same as in Fig.\ \ref{figGav_50}. Furthermore we see many rapid
oscillations with a small amplitude which were absent in Fig.\
\ref{figGav_50}. The reason is that $\bra\phi^2\ket$ is ultraviolet finite
and therefore dominated by the low-momentum modes, while $\bra \pi^2\ket$
is sensitive to all frequencies up to the lattice cutoff.

\begin{figure}
\centerline{\psfig{figure=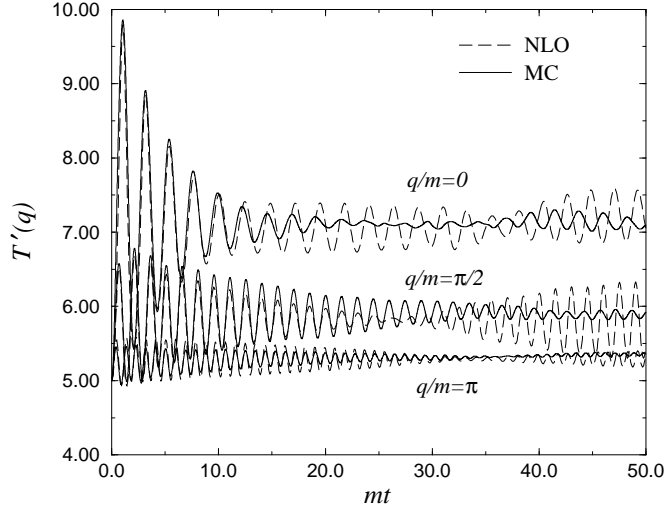,height=7cm}}
\caption{Time evolution of the mode-temperature $T'(q,t) = 
G'_{\pi\pi}(q,t)$ for modes $q/m=0, \pi/2, \pi$. The initial value is
$T_0'=5$ for all $q$. The LO result is not shown for clarity. 
}
\label{figTthreeqs}
\end{figure}

Fig.\ \ref{figTav_50} gives the impression that the system establishes a
new temperature $T'\approx 5.32 \neq T_0' = 5$ rather quickly. However, in
order to be in thermal equilibrium, all momentum modes should have the
same temperature.  In Fig.\ \ref{figTthreeqs} the effective temperature
$T'(q,t)$ for three momentum modes is shown. Perhaps surprisingly, we see
that for each $q$ $T'(q,t)$ oscillates around a different value.
The mean values appear rather stable and do not seem to
approach each other. This resembles the nonthermal fixed points discussed
in \cite{Bonini:1999dh} for the NLO equations.

In fact, it turns out that this behaviour can be understood with
satisfactory accuracy already in terms of fixed or
stationary points of the evolution equations in the Hartree approximation.
The relations between the two-point functions at a fixed point (denoted
with a star) are readily determined from the Hartree equations 
(\ref{eqLO}), and read
\be
\label{eqfixed}
G^*_{\pi\pi}(q) =  \bar\om_q^{*2} G^*_{\phi\phi}(q),\;\;\;\; 
G^*_{\pi\phi}(q) = 0,
\ee
with
\be
\bar\om_q^{*2}=\om_q^2+\frac{3}{2}\lambda\bra\phi^2\ket^*.
\ee 
The first equation shows an expected relation between $\pi\pi$ and
$\phi\phi$ two-point functions, the second expression confirms that the
time-reflection odd two-point function has to vanish at a stationary 
point.

The fixed-point structure of the Hartree equations by itself does not yet 
constrain the allowed fixed-point solutions completely. However, we can
supplement the set of equations (\ref{eqfixed}) with the nontrivial 
combinations $\alpha^2(q)$, given in Eq.\ (\ref{eqalpha}), that
are exactly conserved for each momentum mode $q$ independently. At a 
fixed point this gives a third relation
\be
\label{eqfixedalpha}
G^*_{\pi\pi}(q)G^*_{\phi\phi}(q) = \alpha^{-2}(q).
\ee 
We recall that $\alpha(q)$ can be determined from the initial ensemble.
Combining (\ref{eqfixed}) and (\ref{eqfixedalpha}) yields the
complete fixed-point solution 
\bea
&&G^*_{\pi\pi}(q) = \frac{\bar\om^*_q}{\alpha(q)},\\
&&G^*_{\phi\phi}(q) = \frac{1}{\bar\om^*_q\alpha(q)}.
\eea
Since $G^*_{\pi\pi}(q)$ is identified with the effective temperature for
a mode $q$, the first equation shows that at a fixed point the system will
generically be nonthermal, since the temperature of a mode depends on its
momentum, and that the nonthermal `temperature profile' follows directly
from the initial ensemble.
The second expression leads to a gap equation, after an integration over
$q$,
\be
\bra \phi^2\ket^* \equiv \int \frac{dq}{2\pi}\,G^*_{\phi\phi}(q) = \int
\frac{dq}{2\pi}\, \frac{1}{\bar\om_q^*\alpha(q)},
\ee
since the right-hand side depends on $\bra \phi^2\ket^*$ via
$\bar\om_q^*$.
We would like to stress again that for an arbitrary initial ensemble the
fixed
point in the Hartree approximation is determined completely and all its
properties can be calculated.

We will now become explicit and specialize to the Gaussian initial
ensemble we consider in this paper. From the initial expectation values
(\ref{eqgaussian}) one finds
\be
\alpha(q) = \frac{\om_q}{T_0},
\ee
so that the gap equation reads
\be
\bra \phi^2\ket^* = T_0\int \frac{dq}{2\pi}\,
\frac{1}{\bar\om^*_q\om_q}.
\ee
It is convenient to introduce the dimensionless parameter $\Delta =
3\lambda\bra \phi^2\ket^*/(2m^2) = \bra\phi'^2\ket^*/2$, so that the
above equation can be written as
\be
\Delta = \frac{T_0'}{2\pi}\frac{1}{\sqrt{1+\Delta}}F(\frac{\pi}{2},
\sqrt{\frac{\Delta}{1+\Delta}}),
\ee 
where $F(\pi/2,k)$ is the (complete) elliptic function of the first kind,
and we recall that $T_0'=3\lambda T_0/m^3$. This gap equation can be
solved numerically. For $T_0'=5$, we find $\bra\phi'^2\ket^* = 2\Delta =
2.06$, which can be compared with the time-averaged value of the LO, NLO,
and MC evolution in Fig.\ \ref{figGav_50}. 

\begin{figure}
\centerline{\psfig{figure=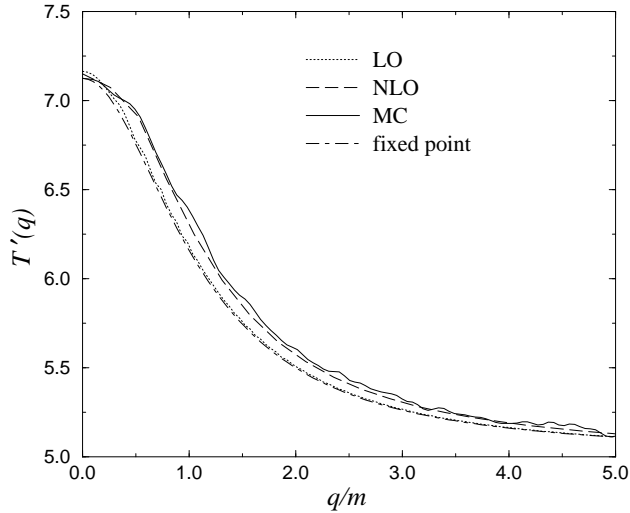,height=7cm}}
\caption{Temperature profile: effective mode-temperature $T'(q)$ versus
momentum $q/m$
after time averaging over the interval
$0<mt<50$, for the LO, NLO, and MC evolution. The fourth line is the
analytic expression at the fixed point of the Hartree equations, with
$\bra \phi'^2\ket^*=2.06$ from the fixed-point gap equation.}
\label{figTprofilezoom}
\end{figure}  

For the effective momentum-dependent temperature at the fixed point we
find 
\be
\label{eqprofile}
T^*(q) = G^*_{\pi\pi}(q) = T_0\frac{\bar\om_q^*}{\om_q} = T_0\left[ 1 +
\frac{3}{2}\frac{\lambda\bra\phi^2\ket^*}{\om_q^2}\right]^{1/2}.
\ee
In order to compare this profile with the numerical results, we calculate
the time average of $T'(q,t)$ between $0<mt<50$ for the LO, NLO, and MC
evolution. The result is shown in Fig.\ \ref{figTprofilezoom}.  We see
that the temperature profile emerging dynamically in the Hartree
approximation is extremely well described by its fixed-point shape
(\ref{eqprofile}).
The result from MC turns out to be remarkably close, implying
that the full nonlinearity only has a small quantitative effect at this
stage, and that direct scattering is not very important. Also, the NLO
profile is close to the MC result, showing that the inclusion of
momentum-dependent four-point functions in the truncation of the effective
action improves the agreement with the full evolution.
Finally, the frequency of oscillation of the individual two-point
functions $G_{\psi\psi'}$ (or $A, B$, and $C$) is approximately
$2\bar\om^*$, which can be seen in Fig.\ \ref{figTthreeqs} for
$G_{\pi\pi}(q,t)$.
We refer to the stage where the dynamics is well described by the
Hartree fixed point as the intermediate-time regime.

At next-to-leading order, the fixed-point structure changes and becomes
much more complicated. From time-reflection symmetry, it is clear that the
two-point function $C(q,t)$, and the four-point couplings
$v(q_1,q_2,q_3;t)$ and $y(q_1,q_2,q_3;t)$ have to vanish at a fixed point.
This implies that also
$\gm(q,t)$ vanishes. However, the other four-point couplings cannot be
zero at a fixed point, which can be seen e.g.\ from the dynamical equation
for $v(q_1,q_2,q_3;t)$ in Appendix \ref{appfourpoint}. Therefore the fixed
point is determined by a set of integral equations. Furthermore, the
relation of the fixed point to the initial ensemble may be
rather complicated.

\section{Late times and thermalization}
\label{seclate}

\begin{figure}
\centerline{\psfig{figure=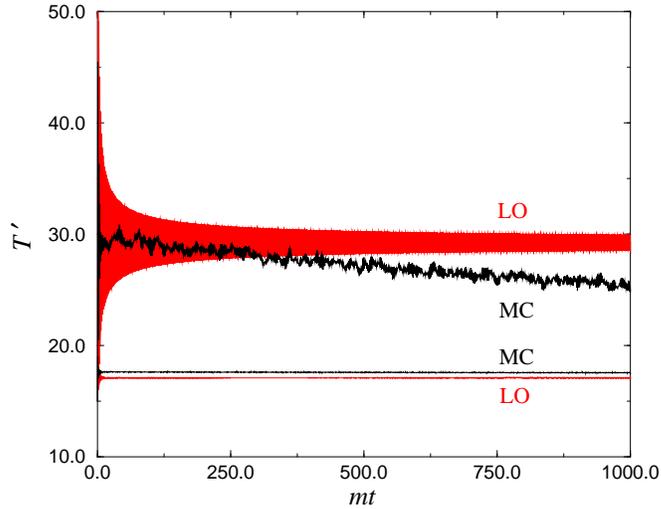,height=7cm}}
\caption{Time evolution of the zero-mode temperature $T'(0,t)$ (two
upper lines) from LO ($N=8192$) and MC ($N=128, N_m=12,\!500$). 
The Hartree fixed-point value is $28.5$. 
The average temperatures over all modes $T'(t)$ (two lower lines) appear
as
straight lines with the MC result slightly above the LO result. The
initial temperature of the Gaussian ensemble is $T'_0=15$. 
}
\label{figrelaxearly2}
\end{figure}  

The results in the previous section show that in the intermediate-time
regime correlation functions appear quasi-stationary but are not thermal.
In particular the time-averaged value of $G_{\pi\pi}(q,t)$ is well
described by the nonthermal profile (\ref{eqprofile}), determined from the
fixed point of the Hartree equations. Since the MC profile is, up to a
small quantitative correction, in agreement with (\ref{eqprofile}) as
well, we infer that this (quasi-)fixed point plays a role also in the full
nonlinear evolution.

The fate of the fixed point can be determined by going to longer times. In
Fig.\ \ref{figrelaxearly2} we show the evolution of the effective
temperature of the zero-momentum mode for LO and MC. We have chosen a
higher value for $T_0'$ than before.  Also shown are the average
temperature over all modes, $T'(t)$.  As was already mentioned in Sec.\
\ref{secearly}, the equilibrium temperature, which we will denote with
$T'$, is established very early in the evolution, so that the lines
presenting $T'(t)$ appear straight. We see that at LO the zero mode
remains oscillating around approximately 29.2. We have calculated the
fixed-point value for $T_0'=15$ and found $T'^*(q=0)=28.5$. The full
nonlinear evolution, on the other hand, shows a decrease towards $T'$: the
approach to thermal equilibrium.  We also
see that the damping at LO is unrelated to the MC result.  It is clear
that whereas the Hartree approximation describes the early and
intermediate regimes qualitatively (or even quantitatively), it is not
able to move away from the fixed point and the approximation breaks down
completely in the late-time regime.

\begin{figure}[t]
\centerline{\psfig{figure=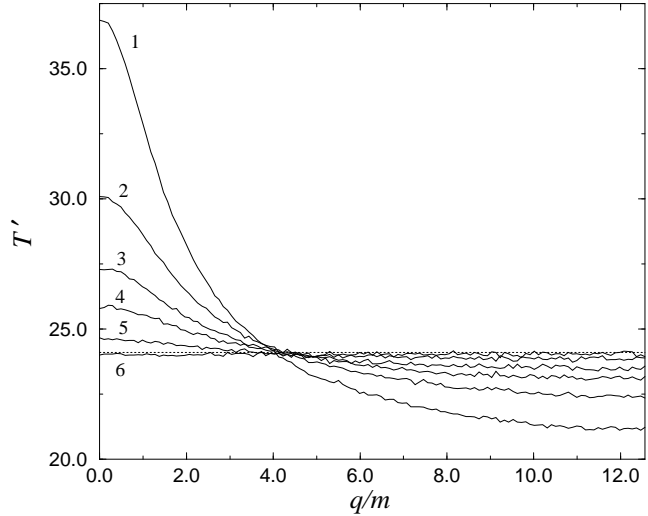,height=7cm}}
\caption{Approach to equilibrium: (time-averaged) snapshots of the
momentum-dependent temperature $T'(q,t)$ for all the modes up to cutoff
$\Lambda/m=4\pi$. The curves represent a time average over
an interval $(t_i-\Delta t, t_i)$, with $m\Delta t = 1500$, and
$mt_1 = 1500$ ($i=1$), 3000 (2), 4500 (3), 
6000 (4), 9000 (5), and 15000 (6).
Line 6 is hardly distinguishable from the dotted horizontal
line, which is the average temperature from all modes ($T'=24.01$).
The parameters are $T_0'=20, N=256, N_m=2,\!500$
(in Figs.\ \ref{figspec}-\ref{figrelaxationrate} MC only).
}
\label{figspec}
\end{figure}  

Unfortunately, for accessible volume sizes the NLO evolution cannot
reach the relevant time scales before becoming unreliable (see the
discussion around Figs.\ \ref{figconvergence}, \ref{figbreaktime}). 
At the largest possible times where the evolution could still be followed,  
we have not been able to see a sign of thermalization.

We continue with MC only. As indicated above, a good observable to follow
during the thermalization stage is the temperature profile
$G_{\pi\pi}(q,t)$, which should become flat ($q$-independent). 
The evolution from the fixed-point profile at
intermediate times to a thermal profile at late times is presented
in Fig.\ \ref{figspec}. We show the time dependence of
$T'(q,t)$ for all modes up to the lattice cutoff, averaged over an
interval $m\Delta t =1500$, for six intervals. 
In the first few intervals, the presence of the nonthermal profile is
still visible. As time goes on, the profile becomes flatter and flatter.
In the last interval shown, between $mt=13500$ and $15000$, the profile
appears $q$-independent and can hardly be distinguished from the straight
line, $T'=24.01$. We see that all momentum modes obtain the same
temperature $T'$ roughly at the same time. For a detailed investigation on
the issue of thermalization in this model concerning other correlation
functions than $G_{\pi\pi}$ we refer to our previous
paper \cite{Aarts:2000mg}. The aspects of thermalization we studied there
are complementary to our findings here (in Ref.\ \cite{Aarts:2000mg} we
focused on the independence of initial conditions, the long-time behaviour
of temporal averages in single `microstates' and other (non-Gaussian)
initial ensembles, and the role of the thermodynamic limit).

\begin{figure}
\centerline{\psfig{figure=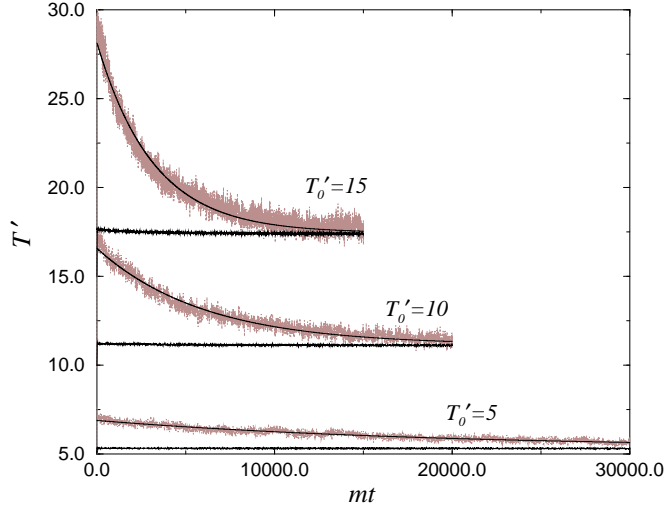,height=7cm}}
\caption{Relaxation of the zero-mode temperature
$T'(0,t)$ to the average temperature for three values of the initial
temperature $T'_0$. The average temperatures $T'(t)$ from all modes
appear as straight lines. Also shown are exponential fits, explained in
the text. The parameters are $N=128, N_m=5,\!000$ each.
}
\label{figrelax1}
\end{figure}

\begin{figure}
\centerline{\psfig{figure=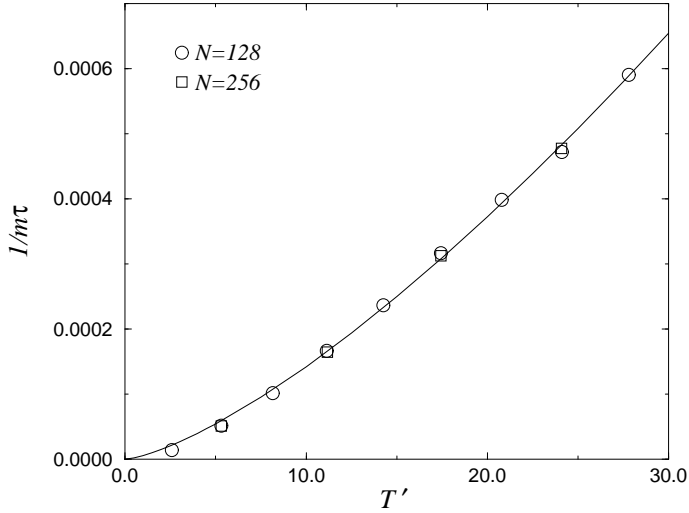,height=7cm}}
\caption{Relaxation rate $1/m\tau$ of the zero-mode $T'(0,t)$ versus the
final equilibrium temperature $T'$.
The initial temperatures are $T_0'=2.5, 5, 7.5, \ldots, 22.5$, for
$N=128$ using $N_m=5,\!000$ initial conditions for each temperature,
and  $T_0'=5, 10, 15, 20$ for $N=256$ ($N_m=2,\!500$). 
Statistical errors are smaller than the symbol sizes.  
The line is a phenomenological power law fit of the $N=128$ data, 
$1/m\tau = 5.8\cdot 10^{-6}\; (T')^{1.39}$.
}
\label{figrelaxationrate}
\end{figure}

To determine the time scale for thermalization, we concentrate on the zero
mode. In Fig.\ \ref{figrelax1} we show the relaxation of $T'(0,t)$ towards
the average temperature $T'(t)$ for three different initial temperatures
$T_0'$.
The numerical data are fitted with an exponential of the form
\be
T'(0,t) = T'\left[1+\kappa e^{-t/\tau}\right].
\ee
The fit is performed over the whole time interval.\footnote{We also
checked for possible power law corrections to the exponential
relaxation, but found no indication for those.}
The resulting relaxation rate $1/m\tau$ is shown in Fig.\
\ref{figrelaxationrate} versus the equilibrium temperature $T'$,
for two system sizes, at fixed lattice spacing $ma=0.25$. No volume
dependence is visible. 

In the remainder of this section, we discuss the thermalization time
scale. In a quantum theory, the relaxation rate is in general related to
the imaginary part of the self-energy \cite{Weldon:1983jn}.  A recent
analysis, applying the relaxation-time approximation to a weakly coupled
quantum scalar field in $3+1$ dimensions, can be found in
\cite{Boyanovsky:1996xx}. At weak coupling, the rate is determined by the
imaginary part of the setting-sun diagram, describing an on-shell
two-to-two scattering process, $\om_\vecp+\om_\veck \to
\om_\vecq+\om_{\vecp-\veck-\vecq}$.  In Appendix \ref{apprelax} we show
how to implement a relaxation-rate (linear response) approximation for a
classical field close to equilibrium. We also give the calculation of the
imaginary part of the classical setting-sun diagram in $1+1$ dimensions,
taken on-shell for arbitrary external spatial momentum.

One should note, however, that in $1+1$ dimensions on-shell two-to-two
scattering is special, since the energy-conservation relation has only two
simple solutions: $q=-k$ and $q=p$. Both solutions give $\om_p+\om_k \to
\om_p+\om_k$, and scattering events of this type do not change the
population numbers for the participating momentum modes.  We find in
Appendix \ref{apprelax} that the naively computed `relaxation rate' is
several orders of magnitude bigger than the thermalization rate observed
in the numerical simulations.

Processes with nontrivial momentum exchange are necessary for
thermalization. This goes beyond two-to-two scattering and occurs only at
higher order. It is possible that a resummation of the self-energy to
two-loop order (i.e.\ including a resummation of the setting-sun diagram)
will determine the relevant time scale. This is suggested by the results
found in \cite{Berges:2000ur}, where (quantum) evolution equations for the
two-point function are solved. Those equations contain a setting-sun type
contribution with fully dressed (in a self-consistent manner) propagators.
This would also cure the collinear singularity that appears in the
lowest-order setting-sun diagram, both in the quantum and in the classical
theory (see Appendix \ref{apprelax}).


\section{Outlook}
\label{secoutlook}         

We have investigated the nonequilibrium time evolution of correlation
functions in field theories.  In order to gain an understanding of the
validity of approximations often used for quantum fields away from
equilibrium, we argued that it is useful to consider the equivalent
problem in classical field theory.  Taking 1+1 dimensional classical
$\phi^4$ theory, discretized on a lattice, as a simple (but often used) 
model, we have implemented a Hartree approximation and a truncation for
equal-time correlation functions containing scattering. The truncated
dynamics was compared with the fully nonlinear results from a numerical
sampling of initial conditions. We believe that our findings are relevant
for the $3+1$ dimensional quantum theory as well.

For the investigated initial Gaussian ensembles the
evolution at early and intermediate times is well reproduced by both
truncations. We found that the Hartree approximation underestimates
damping. This difference becomes more pronounced at larger coupling.
The inclusion of scattering leads to a quantitatively better
agreement with the full numerical evolution.

There is an intermediate-time regime in which correlation functions have
nonthermal values. The characteristic behaviour in this regime can be
understood from the presence of a nonthermal quasi-stationary point, which
is close to a fixed point in the Hartree approximation.
As a consequence of (infinitely many) conserved correlation functions,
the Hartree approximation cannot move away from this fixed point and is 
therefore unable to describe thermalization.

The quartic approximation to the equal-time effective action (NLO) suffers
from a different problem. For a finite system ($N$ finite) the initially
very successful description of the evolution moves away from both the
infinite-volume evolution at NLO and the numerical results at some time
$t_N$. For $t>t_N$ fluctuations grow that are related to the truncation
and not to the physical system. We find that $t_N$ is proportional to $N$,
such that the problem may disappear for $N\rightarrow \infty$. In practice
this is of little help, since only finite $N$ can be investigated
numerically.  At a time $t_{\rm u} \gg t_N$ the evolution becomes
uncontrolled and cannot be followed numerically. For accessible values of
$N$, both $t_N$ and $t_{\rm u}$ are found to be (much) smaller than the
typical relaxation time for thermalization. For this reason, we do
not know whether NLO is {\em in principle} able to describe
thermalization or not.

Since the late-time regime can certainly not be described by the Hartree
approximation, and the quartic approximation including scattering becomes
unreliable on the relevant large time scales, we conclude that the regime
of thermalization is still unsolved in the approach using
evolution equations for equal-time correlation functions.

Concerning the late-time regime, promising results for translationally
invariant ensembles have been obtained recently using an unequal-time
formulation that is time-nonlocal \cite{Berges:2000ur}. It would be
interesting to implement the method employed there in a classical theory
and carry out a similar comparison as we did in this paper for the
equal-time formulations. An open question in that respect is whether a
successful truncation can be found in an equal-time formalism as well or
if the nonlocality is crucial for an analytical description of the
thermalization regime.

\vspace{5mm}

{\bf Acknowledgements}

\noindent{G.A.\ would like to thank Jan Smit for useful discussions. This
work was supported by the TMR network {\em Finite
Temperature Phase Transitions in Particle Physics}, EU contract no.\
FMRX-CT97-0122.}

\renewcommand{\thesection}{\Alph{section}}
\setcounter{section}{0}

\section{Symmetry at leading order}
\label{app}

The Hartree equations conserve the combination $\alpha(q)$, defined
in Eq.\ (\ref{eqalpha}), for each $q$. 
This can be understood as follows \cite{Cooper:1997ii}.  The effective
equations can be rewritten by introducing a set of complex variables
$\xi(q,t)$ as
\bea
\nn
&&G_{\phi\phi}(q,t) =  \xi^*(q,t)\xi(q,t),\\
\label{eqxi}
&&G_{\pi\phi}(q,t) =  \half\left[\dot\xi^*(q,t)\xi(q,t) 
+ \xi^*(q,t)\dot\xi(q,t)\right],\\
\nn
&&G_{\pi\pi}(q,t) = \dot\xi^*(q,t)\dot\xi(q,t).
\eea 
In terms of these the evolution equations (\ref{eqLO}) read
\be 
\ddot\xi(q,t) = -\left(\om_q^2+\frac{3}{2}\lambda\int\frac{dp}{2\pi}\,
|\xi(p,t)|^2\right)\xi(q,t).
\ee
This equation can be derived from an effective Lagrange density
\be
{\cal L}_{\rm eff} = 
\int\frac{dq}{2\pi}\,\left[ |\dot\xi(q,t)|^2 - 
\left(\om_q^2+\frac{3}{4}\lambda\int\frac{dp}{2\pi}\,
|\xi(p,t)|^2\right)
|\xi(q,t)|^2\right],
\ee
which has a global (i.e.\ time independent) symmetry $\xi(q,t)\to
\exp[i\theta(q)]\xi(q,t)$ for each $q$. The corresponding conserved
charge reads
\be
Q(q) = i\left[\xi^*(q,t)\dot\xi(q,t) - \dot\xi^*(q,t)\xi(q,t)\right].
\ee
This charge is in fact the conserved quantity and $Q(q)=2\alpha^{-1}(q)$.

\section{Evolution of the four-point couplings}
\label{appfourpoint}

In this Appendix we list the equations that determine the time evolution
of the four-point couplings. Again, these equations are slightly simpler 
than the ones that can be obtained from setting $N_f=1$ in the equations
for the $O(N_f)$ model \cite{Bonini:1999dh}.

The evolution equations read:
\bean
&& \partial_t u(\1,\2,\3) = \big[\tilde\omega^2_{q_1}v(\1,\2,\3) +
     4\lambda C(\1) - 4\lambda C(\2)S_1(\1+\2,\3)\big]_{SYM},
\\
&& \partial_t v(q_1,q_2,q_3) = \big[2\tilde\omega^2_{q_2}w(\1,\2,\3)
   -4u(\1,\2,\3) + 4\lambda B(\1) 
\\
&&\;\;\;\;\;\;\;\;\;\;\;\;\;\;\;\;\;\;\;\;\;\;\;\;\;
-\gamma(\1) v(\1,\2,\3)
- 4\lambda \{B(\1)S_1(\1+\2,\3) 
\\
&&\;\;\;\;\;\;\;\;\;\;\;\;\;\;\;\;\;\;\;\;\;\;\;\;\;
+ C(\4)S_3(\1+\2,\1) 
+ 2C(\2)S_5(\2+\3,\1)\}\big]_{SYM},
\\  
&& \partial_t w(q_1,q_2,q_3) = \big[3\tilde\omega^2_{q_3}y(\1,\2,\3)
-v(\1,\2,\3) -2v(\2,\4,\1) 
\\
&&\;\;\;\;\;\;\;\;\;\;\;\;\;\;\;\;\;\;\;\;\;\;\;\;\;
- \{\gamma(\1)+\gamma(\2)\}w(\1,\2,\3)
-\lambda \{ C(\3)S_4(\1,\2) 
\\
&&\;\;\;\;\;\;\;\;\;\;\;\;\;\;\;\;\;\;\;\;\;\;\;\;\;
+ 8B(\2)S_5(\2+\3,\1)\}
 -4\lambda B(\1)S_3(-\1-\3,\2)\big]_{SYM},
\nonumber\\
&& \partial_t y(q_1,q_2,q_3) = \big[ 4\tilde\omega^2_{q_4}z(\1,\2,\3) -
2w(\1,\2,\3) 
\nonumber\\
&&\;\;\;\;\;\;\;\;\;\;\;\;\;\;\;\;\;\;\;\;\;\;\;\;
-[\gamma(\1)+\gamma(\2)+\gamma(\3)] y(\1,\2,\3)
-\lambda B(\3)S_4(\1,\2)\big]_{SYM}
\nonumber\\
&& \partial_t z(q_1,q_2,q_3) = -\big[ 
y(\1,\2,\3) + 4\gamma(\1)z(\1,\2,\3)\big]_{SYM},
\eean
where the subscript {\scriptsize $SYM$} implies symmetrization with
respect to the
appropriate permutations of $\1$, $\2$, $\3$, and $\4=-(\1+\2+\3)$. Here
we have used the one-loop integrals
\bean
&&
S_1(\1,\2) = \frac{3}{4}\int_{p_1,p_2}2\pi\delta(p_2-p_1-q_1)\, 
G_{\phi\phi}(p_1)G_{\phi\phi}(p_2) \big[6u(p_2,-p_1,q_2) 
\\
&&\;\;\;\;\;\;\;\;\;\;\;\;\;\;\;\;\;\;
 - 3c(p_1)v(-p_1,p_2,q_2)  +c(p_1)c(p_2)w(-p_1,p_2,q_2)\big],
\\
&&
S_3(\1,\2) = \frac{1}{4}\int_{p_1,p_2}2\pi\delta(p_2-p_1+q_1)\, 
G_{\phi\phi}(p_1)G_{\phi\phi}(p_2)
\big[7v(q_2,-p_1,p_2)
 \\
&&\;\;\;\;\;\;\;\;\;\;\;\;\;\;\;\;\;\;
-8c(p_2)w(p_2,q_2,-p_1) + 7c(p_1)c(p_2)y(p_2,-p_1,q_2) \big],
\\
&& S_4(\1,\2) = 3\int_{p_1,p_2}2\pi\delta(p_2-p_1+q_1+q_2)\,
G_{\phi\phi}(p_1)G_{\phi\phi}(p_2)
\big[w(\1,\2,p_2) 
\\
&&\;\;\;\;\;\;\;\;\;\;\;\;\;\;\;\;\;\;
-3c(p_1)y(q_1,q_2,-p_1) + 6c(p_1)c(p_2)z(p_2,-p_1,q_1) \big],
\\
&&
S_5(\1,\2) = \frac{1}{4}\int_{p_1,p_2}2\pi\delta(p_2+p_1+q_1)\, 
G_{\phi\phi}(p_1)G_{\phi\phi}(p_2)
\big[ v(q_2,-p_1,-p_2) \\
&&\;\;\;\;\;\;\;\;\;\;\;\;\;\;\;\;\;\;
-2c(p_1)w(-p_1,q_2,-p_2) + c(p_1)c(-p_2)y(q_2,-p_1,-p_2) \big].
\eean

\section{Relaxation-time approximation}
\label{apprelax}

For a quantum field slightly away from equilibrium the relation
between the imaginary part of the self-energy and the thermalization rate
for the single-particle distribution function has been pointed out long
ago by Weldon \cite{Weldon:1983jn}. A more recent detailed analysis can be
found in \cite{Boyanovsky:1996xx} for a scalar field in $3+1$
dimensions. Here we briefly outline the arguments of
\cite{Boyanovsky:1996xx} in $d+1$ dimensions and then show how to adapt
those for the {\em classical} theory we consider here. This
analysis is valid for a weakly coupled plasma, close to equilibrium.

A dynamical equation for the (quasi-)particle distribution $n(\vecp,t)$, 
describing the relaxation towards the Bose distribution $n_{\rm
B}(\om_\vecp) = 1/[\exp(\om_\vecp/T)-1]$, can be obtained perturbatively,
using the Heisenberg equations of motion and resumming hard thermal loops
if necessary. In the relaxation-time approximation, the distribution
function for a momentum mode $\vecp$ is written as
\be
 n(\vecp,t) = n_{\rm B}(\om_\vecp) +\delta  n(\vecp,t).
\ee
All other modes $\vecq\neq \vecp$ are assumed to be in equilibrium.
Skipping many steps \cite{Boyanovsky:1996xx}, the
resulting linearized equation is
\be
\partial_t\delta  n(\vecp,t) = -\Gamma(\vecp)\delta  n(\vecp,t),
\ee
with the relaxation rate
\be
\Gamma(\vecp) = -\frac{\mbox{Im}\, \Sigma(\om_\vecp,\vecp)}{\om_\vecp}.
\ee
This rate is twice the plasmon damping rate $\gm(\vecp)$.
For a weakly coupled scalar field with a $\phi^4$ interaction in $3+1$
dimensions, $\Gamma(\vecp)$ is determined by the imaginary part of the
setting-sun diagram.

For the classical theory our interest is in the evolution of the
momentum-dependent `temperature' $T(\vecp,t) = G_{\pi\pi}(\vecp,t)$. We
may obtain this correlation function from the classical unequal-time
two-point function $S(\vecx-\vecy;t_1,t_2) =
\bra\phi(\vecx,t_1)\phi(\vecy,t_2)\ket$ by a spatial Fourier transform as
\be
\label{eqpert}
T(\vecp,t) =  
\partial_{t_1}\partial_{t_2}S(\vecp;t_1,t_2)\Big|_{t_1=t_2=t}.
\ee
A calculation of $S(\vecp;t_1,t_2)$ using classical perturbation theory to
second order in the coupling constant can be found in \cite{Aarts:1997qi} 
for $d=3$ (see also \cite{Buchmuller:1997yw}).  
The resulting expressions are similar to the quantum ones. The most
important change is that the Bose distributions are replaced by classical
distribution functions
\be
n_{\rm cl}(\om_\vecp) = \frac{T}{\om_\vecp}.
\ee
The relevant second-order contribution reads \cite{Aarts:1997qi}
\bean
S_2(\vecp;t_1, t_2) &=&
-\int_0^\infty dt' dt''\,
G^R_0(\vecp,t_1-t')\Sigma_{R,\rm cl}(\vecp,t'-t'')S_0(\vecp,t''-t_2)\\
&& -\int_0^\infty dt' dt''\,
S_0(\vecp,t_1-t')\Sigma_{A,\rm cl}(\vecp,t'-t'')G^A_0(\vecp,t''-t_2)\\
&& -\int_0^\infty dt' dt''\,
G^R_0(\vecp,t_1-t')\Sigma_{F,\rm cl}(\vecp,t'-t'')G^A_0(\vecp,t''-t_2),
\eean
where 
\be
S_0(\vecp,t) = n_{\rm cl}(\om_\vecp)\frac{\cos \om_\vecp t}{\om_\vecp}
\ee
is the free two-point function. The free retarded Green function reads 
\be
G^R_0(\vecp,t) = \theta(t)\frac{\sin \om_\vecp t}{\om_\vecp} =
G^A_0(\vecp,-t),
\ee
and the self-energy corrections are, in $d+1$ dimensions,
\bean
&&\Sigma_{R,\rm cl}(\vecp,t) =
-\frac{9\lambda^2}{2}
\int_{\veck,\vecq}S_0(\veck,t)S_0(\vecq,t)G^R_0(\vecp-\veck-\vecq,t) 
= \Sigma_{A,\rm cl}(\vecp,-t),\\
&&\Sigma_{F,\rm cl}(\vecp,t) = -\frac{9\lambda^2}{6}\int_{\veck,\vecq}
S_0(\veck,t)S_0(\vecq,t)S_0(\vecp-\veck-\vecq,t),
\eean        
with
\be
\int_{\veck} = \int \frac{d^dk}{(2\pi)^d}.
\ee
After performing all the time integrals and taking the derivatives as
in Eq.\ ({\ref{eqpert}), one may follow the arguments
given in \cite{Boyanovsky:1996xx} for the
quantum theory to find the evolution for $\delta T(\vecp,t)\equiv
T(\vecp,t) -T$ in the relaxation-time approximation.  The result is
\be
\partial_t\delta T(\vecp,t) = -\Gamma_{\rm cl}(\vecp)\delta T(\vecp,t),
\ee
with
\be
\Gamma_{\rm cl}(\vecp) = -\frac{\mbox{Im}\,
\Sigma_{R,\rm cl}(\om_\vecp,\vecp)}{\om_\vecp}.
\ee
On-shell, only the two-to-two scattering contribution in the retarded
self-energy is kinematically allowed. This can be written as
\cite{Wang:1996qf,Aarts:1997kp}
\bean
&&\mbox{Im}\, \Sigma_{R,\rm cl}(\om_\vecp,\vecp) =  \\
&&\;\;\;\; -\frac{9\lambda^2}{4}\frac{\om_\vecp}{T}  
\int d\Phi_{123}(\vecp)\,
2\pi\delta(\om_\vecp +\om_\veck-\om_\vecq-\om_\vecr)
n_{\rm cl}(\om_\veck)n_{\rm cl}(\om_\vecq)n_{\rm cl}(\om_\vecr),
\eean
where
\be
d\Phi_{123}(\vecp) = 
\frac{d^dk}{(2\pi)^d2\om_\veck}
\frac{d^dq}{(2\pi)^d2\om_\vecq}
\frac{d^dr}{(2\pi)^d2\om_\vecr}
(2\pi)^d\delta(\vecp-\veck-\vecq-\vecr).
\ee

Specializing to $d=1$, we find
\be
\Gm_{\rm cl}(p) = \frac{9\lambda^2T^2}{64\pi}I(p),
\ee   
with
\be 
I(p) = 
\int_{-\infty}^{\infty}dk\int_{-\infty}^{\infty}dq\,
\frac{\delta(\om_p+\om_k-\om_q-\om_{p-k-q})}
{\om_k^2\om_q^2\om_{p-k-q}^2}.      
\ee
Note that the momentum integrals are ultraviolet finite. The integral is
invariant under $p\to -p$, so we may restrict ourselves to $p\geq 0$. The
$q$-integral can be performed using the delta function, which has support
at two separated points only, $q=p$ and $q=-k$. The result is
\bea
\nn
I(p) &=& 
\frac{2}{\om_p}\int_{-\infty}^{\infty} 
dk\, \frac{1}{\om_k^3}\frac{1}{|k\om_p+p\om_k|}\\
&=& 
-\frac{2}{\om_p}\int_{-\infty}^{-p}
dk\,
\frac{1}{\om_k^3}\frac{k\om_p-p\om_k}{\om_k^2-\om_p^2}
+\frac{2}{\om_p}\int_{-p}^{\infty}
dk\,
\frac{1}{\om_k^3}\frac{k\om_p-p\om_k}{\om_k^2-\om_p^2}.
\eea
The remaining integrals contain a collinear singularity when $k\to
-p$, leading to a logarithmic divergence. This singularity is present in
the quantum self-energy as well. We regulate this in an ad-hoc
manner by modifying the integration boundaries to $-p\pm \mu$, with
$\mu\ll m$ a small cutoff. 

The integrals are straightforward
using partial fractioning, and the final result is
\be
\Gamma_{\rm cl}(p) =  
\frac{9\lambda^2T^2}{16\pi m^2}\frac{1}{\om_p^3}
\left[-1+\frac{p}{m}\arctan\frac{p}{m} + \ln
\frac{2\om_p}{\mu}\right].
\ee
The rate is positive for sufficiently small $\mu$ ($\mu/m \lesssim
0.7$). 
Two limiting cases are 
\be
\Gamma_{\rm cl}(0)  = \frac{9\lambda^2T^2}{16\pi m^5}\ln
\frac{2m}{\mu},\;\;\;\;\;\;
\Gamma_{\rm cl}(p\gg m) = \frac{9\lambda^2T^2}{32\pi m^3p^2}.
\ee
Up to now we have silently neglected the corrections to the mass parameter
due to interactions. If we denote the resummed mass parameter with $M$,
$m$ has to replaced by $M$ in the expressions above.

Let us conclude by noting that loosely identifying $\mu$ with
$\Gm_{\rm cl}$ itself gives a relaxation rate $\Gamma_{\rm cl}(0)$ which
is at least two orders of magnitude too big, when compared to the
numerical results for the thermalization rate $1/\tau$.


\begin{thebibliography}{10}

\bibitem{Kofman:1997yn}
L.~Kofman, A.~Linde, and A.~A. Starobinsky, 
Phys. Rev. {\bf D56} (1997) 3258
[hep-ph/9704452].

\bibitem{Boyanovsky:1996sv}
D.~Boyanovsky, H.~J. de~Vega, and R.~Holman, {\em Erice lectures on
  inflationary reheating}, in {\em International {S}chool of {A}strophysics,
  {D}. {C}halonge: 5th {C}ourse: {C}urrent {T}opics in {A}strofundamental
  {P}hysics},
hep-ph/9701304.

\bibitem{Cooper:1994hr}
F.~Cooper, S.~Habib, Y.~Kluger, E.~Mottola, J.~P. Paz, and P.~R. Anderson, 
Phys. Rev. {\bf  D50} (1994) 2848 
[hep-ph/9405352].

\bibitem{Cooper:1997ii}
F.~Cooper, S.~Habib, Y.~Kluger, and E.~Mottola, 
Phys. Rev. {\bf D55} (1997) 6471 
[hep-ph/9610345].

\bibitem{Boyanovsky:1997cr}
D.~Boyanovsky, D.~Cormier, H.J.~de Vega, R.~Holman, A.~Singh, and 
M.~Srednicki, Phys. Rev. {\bf D56} (1997) 1939
[hep-ph/9703327];
D.~Boyanovsky, D.~Cormier, H.J.~de Vega, R.~Holman, and S.P.~Kumar, 
{\em ibid.\ }{\bf D57} (1998) 2166
[hep-ph/9709232].

\bibitem{Baacke:1997se}
J.~Baacke, K.~Heitmann, and C.~P{\"a}tzold, 
Phys. Rev. {\bf D55} (1997) 2320 [hep-th/9608006];
{\em ibid.\ }{\bf D58} (1998) 125013 [hep-ph/9806205]

\bibitem{Destri:1999hd}
C.~Destri and E.~Manfredini, 
Phys. Rev. {\bf D62} (2000) 025007 [hep-ph/0001177].
 
\bibitem{Boyanovsky:1996zy}
D.~Boyanovsky, M.~D'Attanasio, H.~J. de~Vega, and R.~Holman, 
Phys. Rev. {\bf D54} (1996) 1748
[hep-ph/9602232];
D.~Boyanovsky, H.~J. de~Vega, R.~Holman, S.~P. Kumar, and R.~D. Pisarski, 
{\em ibid.\ }{\bf D58} (1998) 125009 [hep-ph/9802370].

\bibitem{Aarts:1998td} G.~Aarts and J.~Smit,
Nucl. Phys. {\bf B555} (1999) 355 
[hep-ph/9812413];
Phys. Rev. {\bf D61} (2000) 025002 
[hep-ph/9906538].

\bibitem{Cheetham:1996nd}
G.~J. Cheetham and E.~J. Copeland, 
Phys. Rev. {\bf D53} (1996) 4125 
[gr-qc/9503043].

\bibitem{Mihaila:1997gb}
B.~Mihaila, J.~F. Dawson, and F.~Cooper, 
Phys. Rev. {\bf D56} (1997) 5400 
[hep-ph/9705354];
B.~Mihaila, T.~Athan, F.~Cooper, J.~Dawson, and S.~Habib, 
hep-ph/0003105;
B.~Mihaila, F.~Cooper, and J.~F. Dawson, 
hep-ph/0006254.

\bibitem{Braghin:1998yw}
F.~L. Braghin, 
Phys. Rev. {\bf D57} (1998) 6317.

\bibitem{Wetterich:1997ap}
C.~Wetterich, 
Phys. Rev. Lett. {\bf 78} (1997) 3598 
[hep-th/9612206].

\bibitem{Bettencourt:1998nf}
L.~M.~A. Bettencourt and C.~Wetterich, 
Phys. Lett. {\bf B430} (1998) 140 
[hep-ph/9712429].

\bibitem{Bettencourt:1998xb}
L.~M.~A. Bettencourt and C.~Wetterich, 
hep-ph/9805360.

\bibitem{Bonini:1999dh}
G.~F. Bonini and C.~Wetterich, 
Phys. Rev. {\bf D60} (1999) 105026 
[hep-ph/9907533].

\bibitem{Ryzhov:2000fy}
A.~V. Ryzhov and L.~G. Yaffe, 
hep-ph/0006333.

\bibitem{Berges:2000ur}
J.~Berges and J.~Cox, 
hep-ph/0006160.

\bibitem{Wetterich:1997rp}
C.~Wetterich, 
Phys. Rev. {\bf E56} (1997) 2687 
[hep-th/9703006].

\bibitem{Aarts:2000mg}
G.~Aarts, G.~F. Bonini, and C.~Wetterich, 
Nucl. Phys. {\bf  B587} (2000) 403
[hep-ph/0003262].
 
\bibitem{Weldon:1983jn}
H.~A. Weldon, 
Phys. Rev. {\bf D28} (1983) 2007.

\bibitem{Boyanovsky:1996xx}
D.~Boyanovsky, I.~D. Lawrie, and D.~S. Lee, 
Phys. Rev. {\bf D54} (1996) 4013 
[hep-ph/9603217].

\bibitem{Aarts:1997qi}
G.~Aarts and J.~Smit, 
Phys. Lett. {\bf B393} (1997) 395 
[hep-ph/9610415].

\bibitem{Buchmuller:1997yw}
W.~Buchm{\"u}ller and A.~Jakov{\'a}c, 
Phys. Lett. {\bf B407} (1997) 39 
[hep-ph/9705452].

\bibitem{Wang:1996qf}
E.~Wang and U.~Heinz, 
Phys. Rev. {\bf D53} (1996) 899 
[hep-ph/9509333].

\bibitem{Aarts:1997kp}
G.~Aarts and J.~Smit, 
Nucl. Phys. {\bf  B511} (1998) 451 
[hep-ph/9707342].

\end{thebibliography}

\end{document}